\documentclass[preprint,12pt]{elsarticle}
\usepackage{graphicx}
\usepackage{pgf}
\usepackage{amsmath}
\usepackage{amssymb}
\usepackage{mathrsfs} % para formato de letra
\usepackage{theorem}

\usepackage{pstricks}
\usepackage[utf8]{inputenc}

\usepackage[normalem]{ulem}

\journal{Draft}

\begin{document}

\begin{frontmatter}

%\runauthor{Antonio Sala}
%\begin{frontmatter}%elsevier
\title{Autonomous Vehicle front steering control computation saving}
\author{Juli\'{a}n Salt Llobregat, Juli\'{a}n Salt Ducajú }
%\thanks{J. Salt was Visiting Scholar at Department of Mechanical Engineering, Universiy of California at % Berkeley. USA }
%A. Sala is with the Systems Engineering Department, Universidad
%Politecnica de Valencia, Cno. Vera, s/n, E-46022 VALENCIA, SPAIN.\
%e-mail: asala@isa.upv,es, julian@isa.upv.es} }
\address{Systems Eng. and Control Dept., Instituto de Automatica e Informatica Industrial, Universitat
Polit\`{e}cnica de Valencia, Cno. Vera, s/n, E-46022 VALENCIA, SPAIN.%\\
%A. Sala is with the Systems Engineering Department, Universitat
%Politecnica de Valencia, Cno. Vera, s/n, E-46022 VALENCIA, SPAIN.
\\
e-mail: jsalt@upv.edu.es}
%\thanks[Someone]{Partially supported by somebody.}
%\date{}
%\maketitle

\begin{abstract}  

For autonomous vehicles lane keeping purposes it is crucial to control the vehicle yaw rate. As it is known a vehicle yaw rate control can be achieved handling the steering angle.  One option is to consider a robust controller and depending of the requirements the synthesis can drive to a high order controller. Nowadays this kind of vehicles needs a networked based control (IVN -Intelligent Vehicle Network-)with a considerable amount of control loops for different vehicle components. Therefore, in this environment the controllers computation saving could be a good option for unload the network and digital processors. That is the main target of this contribution; in order to accomplish this goal a interlacing implementation technique is considered. Results in a real path tracking illustrates viability of this procedure. 

\end{abstract}

%\end{frontmatter}
%\title{Frequency-response of dual-rate systems}
%\maketitle
\end{frontmatter}

\section{Motivation}

Autonomous vehicles have been gaining popularity for some time now because they are set to change the paradigm of transportation systems. Specifically, in autonomous ground vehicles (AGVs), a key issue in achieving automated driving is obtaining controllers that allow a pre-established route to be followed. In this type of vehicle, longitudinal and lateral control are considered. Lateral control deals with lane keeping, which is the objective of this study. For describing AGV’s lateral dynamics a Linear Parameter Varying (LPV) model is usually obtained. The parametric uncertainty depends mainly of variables like vehicle longitudinal speed and lateral acceleration if there are not high lateral tire forces involved.

There are different models \cite{ducaju2020application} and different control techniques applied to this problem \cite{artunedo2024lateral,salt2021autonomous,you2006active}. Due to the model parametric uncertainty, robust controllers can be considered and $\mu$ synthesis control is one of the solutions. This kind of regulators usually have poles at different frequencies. The big problem is that these controllers with a design procedure involving different weights in loop control specifications leads to high order functions. A vehicle good performance requires that the electronic control units and the communications networks (CAN, LIN,...) do not be overloaded. It is what is called a Resource Constrained Environment with limited processing capacity that requires a reduction of computational complexity.
For this intention, an interlacing controller implementation procedure \cite{bhattacharya2009control,wu2004performance,ding2006multirate,salt2014hard} is going to be assumed. 

Basically, this technique consists of applying the contributions of the fast poles at control instants and deferring the contributions of the slow poles by N sampling instants. This basic idea is depicted in figure \ref{inter_basic}. 

In section \ref{seccion_control} is introduced the control problem to be solved, the specific model that has been considered and the controller design procedure although this is not the goal of this contribution; the main objective is to apply the interlacing procedure to a high order controller in order to get a computationally saving. Then a brief introduction to interlacing will be exposed in section \ref{sec_interlac}. Section \ref{lifting_interlac} is devoted to introduce the formal discrete lifting modeling procedure for these interlacing cases. Then, in section \ref{AV_application} a specific application to a real autonomous vehicle will be considered comparing the results (path following) with and without computationally saving. Finally conclusions will close the contribution.

\section{Control problem} \label{seccion_control}
\begin{figure} \centering
\includegraphics[height=5cm, width=12cm]{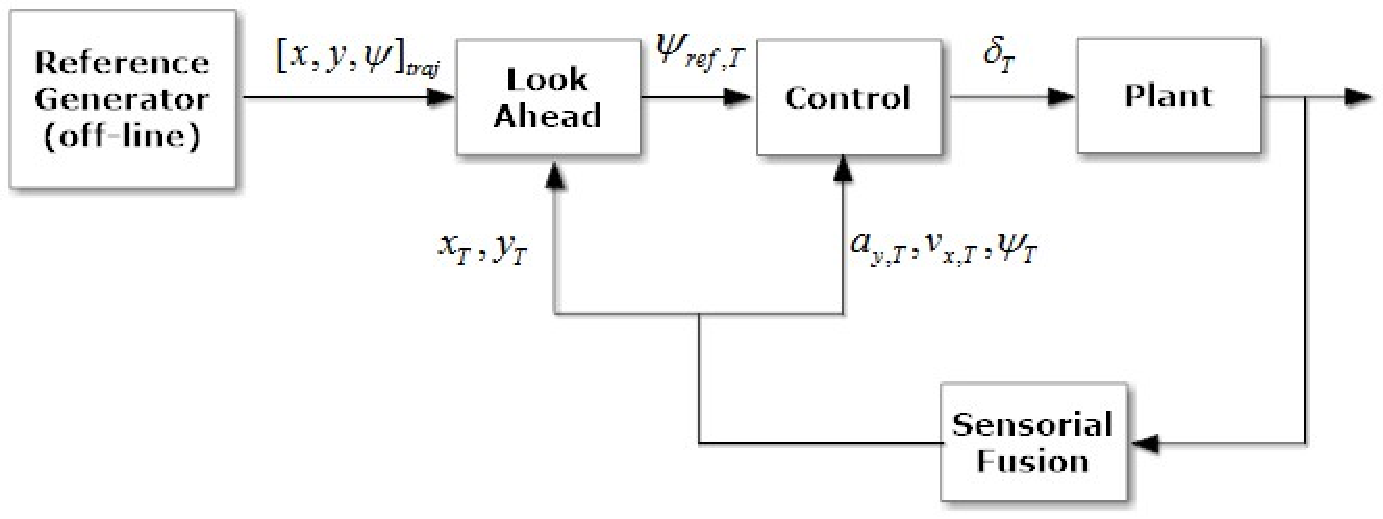}
\caption {Proposed Loop control} \label{control_prop}
\end{figure}

There are diverse control laws devoted to vehicle lane-keeping, commonly called
steering controllers.
%In this section, two widely used methods with some variations are
%considered: the Inverse Kinematic Bicycle model (IKIBI) and the Linear Parameter Varying-
%Model Predictive Control (LPV-MPC).
In both cases, the purpose is to use the steering front wheels’ angle $\delta$ as the control
action in order to follow the desired path. The complete path, $[X,Y,\psi]_{ref}$ is planned
offline, and depending on the controller election, the next yaw rate, $\dot{\psi}_{ref}$, or yaw position
goal, ${\psi}_{ref}$ , is delivered by a pure pursuit procedure with a coherent look-ahead distance
L \cite{coulter1992implementation, kuwata2009real,ljungqvist2015motion,buehler2009darpa}. Figure \ref{control_prop} shows a schematic view of this process.
\subsection{Car lateral dynamics modelling}
There are different ways to expose the AGV model \cite{ducaju2020application}. In our case, the model by \cite{rill2005vehicle} will be considered. The purpose is to obtain a relation between the yaw angle an the steering assuming the vehicle's dynamics in the horizontal plane by neglecting the pitch and roll motion and small slip angle.
As usual, from lateral force and moment balance:
\begin{equation}
ma_y=F_{y,f}cos(\delta)+F_{y,r} \\
I_zr=l_fF_{y,f}cos(\delta)+l_rF_{y,r}
\end{equation}
where $m$, the vehicle body mass, $l_f$ and $l_r$, the distance of the front and rear wheels, respectively, from the normal
projection point of vehicle’s center of gravity onto the common axle plane,  $I_z$ the vehicle body moment of inertia about the vehicle-fixed z-axis. Furthermore, subscripts $f$ and $r$ refer to the front and rear axles, respectively. Another important parameter is $C_{\alpha,fr}$ is the cornering stiffness. This constant represents a linear approximation for the relationship between the slip angle, $\alpha$, and the lateral force, $Fy$, that is:
\begin{equation}
F_{y,f}=C_{\alpha,f} \alpha_f \\
F_{y,r}=C_{\alpha,r} \alpha_r
\end{equation}
The sideslip angle could be defined as:
\begin{equation}
\beta=arctan(\frac{u_y}{u_x})
\end{equation}
being $u_y$ and $u_x$ the longitudinal an lateral velocities.

As it is known \cite{zhang2008development}, it can be expressed:
$$
\begin{bmatrix}
\dot{\beta} \\
\dot{\psi}
\end{bmatrix}
=
\begin{bmatrix}
\frac{-C_{\alpha,f}-C_{\alpha,r}}{mV_x} & -1+\frac{l_r C_{\alpha,r}-l_f C_{\alpha,f}}{mV_x^2} \\
\frac{l_r C_{\alpha,r}-l_f C_{\alpha,f}}{I_z} & \frac{-l_r^2 C_{\alpha,r}-l_f^2 C_{\alpha,f}}{I_zV_x}
\end{bmatrix}
\begin{bmatrix}
\beta \\
\psi
\end{bmatrix}
+
\begin{bmatrix}
\frac{C_{\alpha,f}}{mV_x} \\
\frac{l_fC_{\alpha,f}}{I_z}
\end{bmatrix}
\begin{bmatrix}
\delta    
\end{bmatrix}
$$
\\

This model leads to a transfer function between yaw rate and steering:
\begin{equation}
\frac{\psi}{\delta}=\frac{a_1s+a_2}{b_1s^2+b_2s+b_3} \label{tf_coche}
\end{equation}
being:
\begin{align}
 a_1 &= mV_x l_f C_{\alpha,f}   \\
 a_2 &=(l_f+l_r)C_{\alpha,f}C_{\alpha,r}\\
 b_1 &=mV_xI_z\\
 b_2 &=I_z(C_{\alpha,f}+C_{\alpha,r})+m(l_f^2 C_{\alpha,f}-l_r^2 C_{\alpha,r}) \\
 b_3 &=\frac{C_{\alpha,f}C_{\alpha,r}}{V_x}(l_f+l_r)^2[1+\frac{mV_x(l_r C_{\alpha,r}-l_f C_{\alpha,f})}{(l_f+l_r)^2C_{\alpha,f}C_{\alpha,r}}  ]\\
\end{align}

In general, as it was said, for describing AGV’s lateral dynamics a Linear Parameter Varying (LPV) model is obtained. The parametric uncertainty depends mainly of variables like vehicle longitudinal speed and lateral acceleration if there are not high lateral tire forces involved. This issue will be considered when discussing vehicle control in section \ref{AV_application}
A continuous controller was designed using \textit{musyn} $Matlab$ command. This $\mu$ synthesis considered frequency weights for sensitivity and control signal amplitude limitation. As mentioned above, it should be noted that the objective of this contribution is not to design a controller, but rather to discuss the implementation of a controller.
%Finally, Table A1 shows the numerical values used for the constants presented earlier.

\section{Interlacing} \label{sec_interlac}

Some contributions were introduced assuming this topic. In \cite{wu2004performance,ding2006multirate,wu2008precision,salt2014hard} the controller interlacing implementation for track-following control for hard disk drive is considered. The performance and aliasing analysis is also developed. In \cite{bhattacharya2009control} a LTI continuous controller is transformed into a multirate controller by discretization of slow and fast modes at different sampling periods (multiples by a factor of $N$ between them). Some different structures for interlacing the slow modes are studied using singular value decomposition. There are other trends using the interlacing for design step planning \cite{jia2008new,lopez2010two} considering interlacing for obtaining the slow part regarding diverse performance criteria.\\
In the current contribution, the starting point is the fast single-rate controller. In \cite{arxiv_interlac} a new treatment of real and complex controller's poles will be introduced and a complete frequency response study was assumed for different blocks order implementation establishing some general rules. In order to analyze properly this kind of structures the discrete lifting modeling must be considered because a dual-rate system appears.
%\section{Interlacing Statement: Slow Modes}
\begin{figure} \centering
\includegraphics[height=9cm, width=12cm]{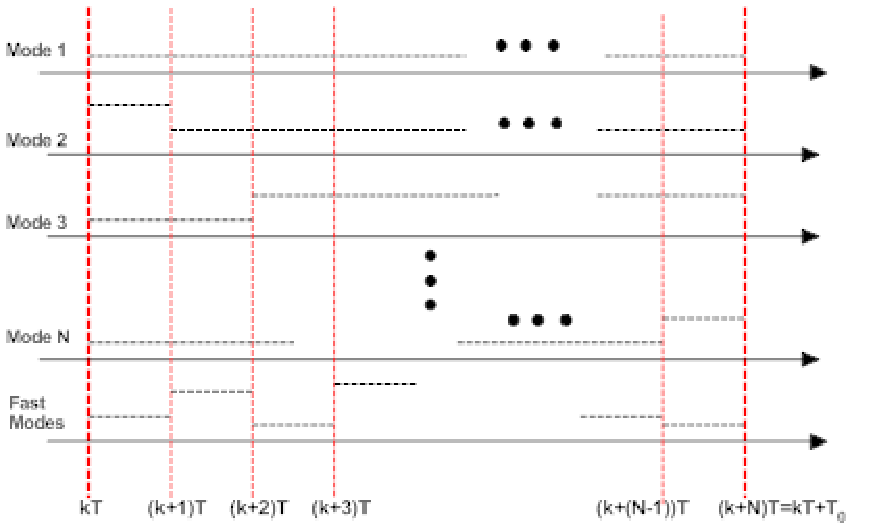}
\caption {Interlacing. Basic Procedure} \label{inter_basic}
\end{figure}

%\subsection{Preliminaries}
%First of all some basic multirate definitions, operations and properties are needed.
%$F^T$ will denote either the Z-transform of the sequence $\{ f(kT) \}$ obtained by sampling the continuous signal $f(t)$ or the sampling rate transformation of a discrete signal $F$. More explanation will be given below.
%$$ F^T(z)= \mathcal{Z} ^T [ \{f(kT)\} ] =\sum_{k=0}^{\infty} f(kT) z^{-k} $$
%in the same way, if the sampling period is $NT$
%$$ F^{NT}(z^N)=\mathcal{Z} ^{NT}[ \{ f(kT) \} ]=\sum_{k=0}^{\infty} f(kNT) z^{-kN} $$
%Now it is defined the upsampling (from $NT$ to $T$) and downsampling (from $T$ to $NT$) transforms:
%$$ [F^{NT}(z^N)]^T=\bar{F}^T(z^N)=\sum_{k=0}^\infty \bar{f}(kT) z^{-kN} =
%\begin{cases}  \bar{f}(kT)=f(kT); \hspace{0.3cm} \forall k=\lambda N \\
%\bar{f}(kT)=0; \hspace{0.3cm} \forall k \neq \lambda N 
%\end{cases} $$

%$$ [F^{T}(z)]^{NT}=\hat{Y}^{NT}(z^N)=\sum_{k=0}^\infty f(kNT) z^{-kN}=F^{NT}(z^N) $$

%Some upsampling-downsamling properties are:
%\begin{equation*}
%\begin{split} 
%[X^T Y^T]^{NT} & \neq [X^T]^{NT}[Y^T]^{NT} \\
%[X^{NT} Y^{NT}]^{T} & = [X^{NT}]^{T}[Y^{NT}]^{T}\\
%[X^T[Y^{NT}]^T]^{NT} & =[X^T]^{NT}Y^{NT}
%\end{split}
%\end{equation*}

\subsection{Problem Statement} \label{Csi}
Given a discrete controller $C(z)$ considering its sampling period what later will be considered fast sampling time, the Interlacing procedure is based on identifying with some criteria slow and fast poles and applying them in different periods. The decomposition can be into series or parallel terms, but the second one is chosen because allows a clearest understanding. Anyhow, the goal is to implement the fast contributions at every fast sampling time and the slow part terms interspersed for equilibrating the computational load in every fast sampling instant. If the slow parts were calculated in slow moments, the time when both contributions coincided would cause an overload on the processor. So, it is decided to apply every single slow contribution in successive fast moments. With this kind of implementation a slow part is applied every slow sampling time. Therefore, this procedure leads to a linear periodically variable system. Figure \ref{inter_basic} helps to understand the intention. Therefore, when a single-rate controller is desired to be implemented with interlacing, some poles must be with different dynamics making sense to this operation. It could be assumed some dynamic areas selected according to some rule for determining what is considered slow or fast; it is usual to consider a slow pole if it is located from to one fifth of Nyquist frequency \cite{wu2008precision}.

A slow part fraction expansion will be needed. The fast discrete controller slow part in parallel terms thinking in an interlacing implementation.
$$C^T_s(z)=\frac{N^T_{cs}(z)}{D^T_{cs}(z)}  $$
where $T$ indicates the discrete transfer function sampling period and $N^T_{cs}(z)$ and $D^T_{cs}(z)$ are polynomials in $z$, where the variable $z$ stands for the LTI z-transform argument at sampling period T, and consequently $z^N$ is related to $NT$. The poles and zeroes of $C^T_s(z)$ are denoted $\alpha_i$ and $\beta_j$ respectively.
The first problem that is found is how to resample a fast sampling period slow pole to a slow sampling period slow pole.

Following \cite{lu1989least} is possible to transform a fast discrete transfer function into a form from which is viable to apply a fast input but every $N$ instants. Specifically, if a polynomial $W^T(z)$ is assumed:
$$\mathcal{W}^T(z)=\displaystyle\prod_{i=1}^{n} \left( z^{N-1}+\alpha_{i,T}z^{N-2}+\hdots +\alpha_{i,T}^{N-1}\right)$$
then:
$$C^T_s(z)=\frac{U^T(z)}{E^T(z)}=\frac{N^T_{cs}(z)}{D^T_{cs}(z)}= \frac{N^T_{cs}(z)\mathcal{W}^T(z)}{D^T_{cs}(z) \mathcal{W}^T(z)}= \frac{\tilde{N}^T_{cs}(z)}{[D^{NT}_{cs}(z^N)]^T} = \frac{U^T(z)}{[E^{NT}(z)]^T} $$

being $E^T(z)$ the input and $U^T(z)$ the output.that is, the input can be a $NT$ period signal and the output will be a $T$ sampling period signal.
If a new polynomial is defined:
$$\mathcal{W}_H^T=\left[ \frac{1-e^{NTs}}{s} \right]^T=\frac{1-z^{-N}}{1-z^{-1}}= 1+z^{-1}+\hdots +z^{N-1} $$
then it is possible to express:
$$
\left[ \frac{U^T(z)}{[E^{NT}(z)]^T} \right]^{NT}=\frac{[U^T(z)]^{NT}}{[[E^{NT}(z)]^T]^{NT}}=\frac{[U^T(z)]^{NT}}{E^{NT}(z)}=\frac{[ \mathcal{W}_H^T \tilde{N^T_{cs}(z)}^T]^{NT}}{[[D^{NT}_{cs}(z^N)]^T]^{NT}}=\frac{N^{NT}_{cs}(z^N)}{D^{NT}_{cs}(z^N)}
$$
that is, a pure slow block, with slow input and output is obtained.

\subsection{Interlacing Implementation Strategies}
The following point is to discuss how can be implemented the slow part of the fast controller. It must be distinguished strategies for input and for output \cite{bhattacharya2009control}.
Regarding the input, the slow terms could assume:
\begin{itemize}
\item I-1. Fast Input
\item I-2. Slow Input
\item I-3 Mean Input 
\end{itemize}
In this contribution it will be used the current fast input in each slow block of fast controller according to its order implementation (see figure \ref{inter_basic}) (I-1). Note that each slow block update is occurring once every $N$ instants. 
In the second case, I-2, all slow blocks are fed by the same slow sampling valid in the $NT$ metaperiod. The third option is to consider the mean of the current and $(N-1)$ fast sampling of the input signal $E^T$ for the slow block input. This last option will not be considered because it is needed some additions and the computation saving is partially lost.
For more details see \cite{arxiv_interlac}.
\\
With respect to the output, it can be observed basically two options. The slow block output is kept during the $NT$ metaperiod and updated when the block is switched on (O-1) or every slow block output is stored and just all slow blocks addition is injected at the end of the metaperiod according to slow sampling times (O-2). O-1 will be noted like fast change or only fast and O-2 like slow change or only slow.
\subsection{Suggestions. Notes}
If it is desired to analyse these structures, it is needed a dual-rate model and one classical method is to consider a discrete lifting \cite{francis1988stability} in order to explain this linear periodically variable system.
If in an initial step a continuous controller was designed, it would be possible to obtain slow and fast single rate controllers. In this case, it is noted that a previous step to analyze if this procedure can be applied is to study if the control loop including the slow controller has a reasonable behavior. In the case that this slow control would result in losing performance or even instability then it would be inadvisable to consider this kind of implementation. As it seems to be a matter of common sense, the implementation of interlacing should be between slow and fast control options.

\section{Interlacing Lifting Application} \label{lifting_interlac}
It is clear that the successive switches of all slow blocks leads to a periodic operation (LPV model) which needs the lifting procedure in order to obtain a LTI function.
In this case a special care must be considered when the discrete lifting modeling is applied on this environment. First of all the complete operation is $NT$ periodic. So, if lifting is wanted to be applied, the blocks order implementation should be studied in one metaperiod. It is going to analyze different options. The problem with $N$ slow poles at fast controller is considered. For all input cases I-{1,2} the problem is similar because a certain signal value arrives at the moment of the switch of a certain slow block. After the current explanation some considerations will be made about the input signal selection. Now, the problem is that for blocks $i=2 \hdots N$ there are $(i-1)T$ sampling periods where the output signal is the same that at preceding metaperiod and $NT-(i-1)T$ outputs computed with the current input signal value. For the first control actions the lifting modeling requires a new state which will be a dummy variable $\chi$ that will represent the control action at the end of the previous metaperiod.
\begin{equation}
\left(
\begin{array}{c}
x\\
\chi \\
\end{array}
\right) _{(k+1)T} =
\left( \begin{array}{lc}
A_i & B_i \\
0 & 0 \\  
\end{array} 
\right) 
\left(
\begin{array}{c}
x\\
\chi \\
\end{array}
\right) _{kT} +
\left(
\begin{array}{c}
0\\
1\\
\end{array}
\right) e_k
\end{equation}
\begin{equation}
u_k = 
\left( \begin{array}{lc}
C_i & D_i \\
\end{array} 
\right) 
\left(
\begin{array}{c}
x\\
\chi \\
\end{array}
\right) _{kT} +
\left(
\begin{array}{c}
0\\
\end{array}
\right) e_k
\end{equation}
being $(A_i,B_i,C_i,D_i)$ the state space matrices of slow block $i$ at sampling period $NT$.
Note that just one dummy variable is needed because the control action follows from just one previous metaperiod.
Usually the lifting procedure uses a quadruple representation packing the matrices from a general state space representation $\textit{A},\textit{B},\textit{C},\textit{D}$:
\begin{equation}
Block_i \equiv \left(
\begin{array}{c|c}
\textit{A} & \textit{B} \\ \hline
\textit{C} & \textit{D} \\
\end{array}
\right)
\end{equation}
As it was said before, the input signal treatment completes the lifting modeling.
The expression for representing the fast input selection into a metaperiod will be:
\begin{equation}
Block_i 
\left(
\begin{array}{c}
0\\
\vdots \\
1 \hspace{0.2cm} \text{file i}\\
\vdots \\
0 \\
\end{array}
\right)^{t}
\end{equation}
that is with a ``1" selecting the input instant order in the metaperiod. Note that in one metaperiod $NT$ there are $N$ values lifted from fast signals at $T$.
Finally the sampling period updating in the metaperiod will be described by means of a similar operation. For instance for block $i=3$ the contribution every fast sampling period into a metaperiod will be:
\begin{equation}
\left(
\begin{array}{c}
1\\
1\\
1\\
0\\
\vdots\\
0 \hspace{0.2cm} \text{file N}\\
\end{array}
\right) block_{\chi_{i}} +
\left(
\begin{array}{c}
0\\
0\\
0\\
1\\
\vdots \\
1 \hspace{0.2cm} \text{file N}\\
\end{array}
\right) block_{i}
\end{equation}
As an example to understand the procedure, it is considered the case of a fast controller with four modes $b_j$ being three of them $b_2,b_3,b_4$ slow ones. It is considered that the implementation is $b_1$ fast and $b_4,b_2,b_3$ with this order. Therefore $N=3$ and the option (I-1,O-1), will require the following open-loop lifting modeling:
%& [1 1 1]'*b4k*[1 0 0]+
%&+[1 0 0]'*b2k*[0 1 0]+[0 1 1]'*b2ak*[0 1 0]+[1 1 0]'*b3k*[0 0 1]+[0 0 1]'*b3ak*[0 0 1]+b1k;
\begin{equation}
\begin{split}
&\left( \begin{array}{ccc} 1 &1 &1 \end{array} \right)' *bk4* \left( \begin{array}{ccc} 1 &0 &0 \end{array} \right)+ \\
&+ \left[ \left( \begin{array}{ccc} 1 &0 &0 \end{array} \right)'*bk2+\left( \begin{array}{ccc} 0 &1 &1 \end{array} \right)'bk2_{\chi}  \right]*\left( \begin{array}{ccc} 0 &1 &0 \end{array} \right)+ \\
&+ \left[ \left( \begin{array}{ccc} 1 &1 &0 \end{array} \right)'*bk3+\left( \begin{array}{ccc} 0 &0 &1 \end{array} \right)' bk3_{\chi}  \right]*\left( \begin{array}{ccc} 0 &0 &1 \end{array} \right) \\
& +bk1
\end{split}
\end{equation}
The block $bk1$ does not need the selector vectors because is used in all fast instants.\\
The case (I-1,O-2) will be modeled by:
\begin{equation}
\begin{split}
&\left( \begin{array}{ccc} 1 &1 &1 \end{array} \right)' *bk4* \left( \begin{array}{ccc} 1 &0 &0 \end{array} \right)+ \\
&+ \left[ \left( \begin{array}{ccc} 1 &0 &0 \end{array} \right)'*bk2+\left( \begin{array}{ccc} 0 &1 &1 \end{array} \right)'bk2_{\chi}  \right]*\left( \begin{array}{ccc} 1 &0 &0 \end{array} \right)+ \\
&+ \left[ \left( \begin{array}{ccc} 1 &1 &0 \end{array} \right)'*bk3+\left( \begin{array}{ccc} 0 &0 &1 \end{array} \right)' bk3_{\chi}  \right]*\left( \begin{array}{ccc} 1 &0 &0 \end{array} \right) \\
& +bk1
\end{split}
\end{equation}

\section{Specific Autonomous Vehicle Application} \label{AV_application}
Now, a specific problem is considered. The study is carried out using the vehicle parameters of a 2017 Lincoln in the figure \ref{coche} on
a circuit path. The sampling period of the simulated discrete-time plant was assumed to be
$T = 0.01$ s, which is the same as the fastest acquisition frequency of sensors installed in the
test-bed vehicle. A controller is designed for the autonomous vehicle with parameters in the table \ref{param_coche1}:
\begin{figure} \centering
\includegraphics[height=5cm, width=9cm]{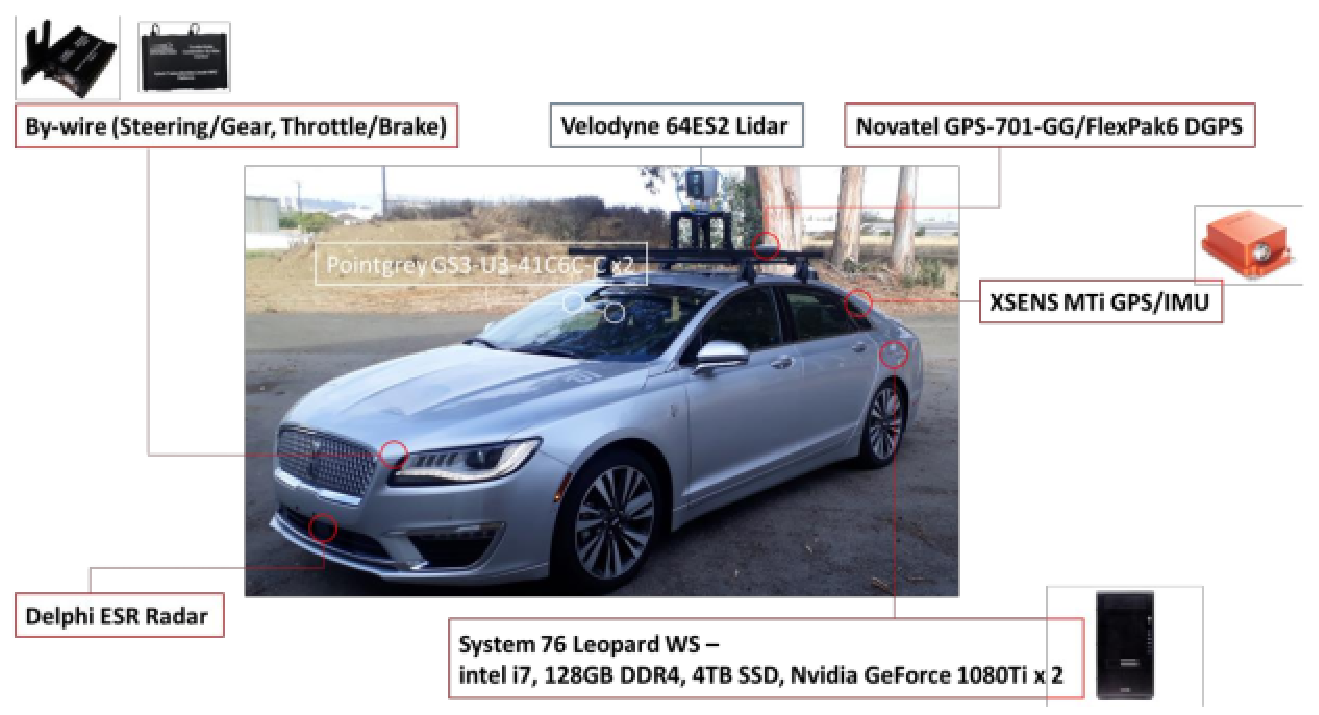}
\caption {Autonomous Vehicle} \label{coche}
\end{figure} 

\begin{table} 
    \centering
    \begin{tabular}{ccc}
    \hline \\
      Constant & Value & Unit  \\
      \hline \\
        m & 1800  & kg\\
       $l_f$ & 1.6 & m\\
       $l_r$ & 1.65 & m\\
      $ C_{\alpha,f}$ & 120 & kN/rad \\
       $C_{\alpha,r}$ &110 & kN/rad\\
       $I_z$ & 3270 & $kg.m^2$ \\
       \hline
    \end{tabular}
    \caption{AGV parameters}
   \label{param_coche1}
\end{table}

In the urban test circuit the longitudinal the velocity varies between 4 and 10 $m/s$. The nominal value was chosen to be 6 $m/s$. In that case, considering (\ref{tf_coche}), it is obtained

\begin{equation}
    \frac{\psi}{\delta}=\frac{5.87s+2.43}{s^2+5.45s+7.21}
\end{equation}

%The control design procedure is a mixed sensitivity $\mathcal{H}^\infty$ based on. 
The continuous $\mu$ synthesis controller obtained was:

\begin{flushleft}
\begin{multline*}
C_0(s)=\frac{2961.5 (s+37.14) (s+3.587) (s+2.616) (s+2.32) (s+2.268) }{(s+1473) (s+37.06) (s+2.531) (s+2.259) (s+1.89) (s+1.535) } 
\\
\times
\frac{(s+2) (s^2 + 12.16s + 81.55)}{(s+0.00796) (s^2 + 12.22s + 81.91)}
\end{multline*}
\end{flushleft}
it is possible to reduce the order using a balanced truncation until sixth order with proper control loop behavior. Then eliminating one dramatically fast pole with gain adjustment, a definitive fifth order controller is obtained:
$$
C(s)=\frac{2.0107 (s+3.937) (s^2 + 5.321s + 7.196) (s^2 + 23.06s + 178.9)}{ (s+3.093) (s+1.523) (s+0.00796) (s^2 + 23.13s + 181.2)}
$$
%Simulations were carried out using the vehicle parameters of a 2017 Lincoln MKZ on
%a circuit path. The sampling period of the simulated discrete-time plant was assumed to be
%$T = 0.01$ s, which is the same as the fastest acquisition frequency of sensors installed in the
%test-bed vehicle.

The discretization for $T=0.01$ s leads to the following discrete controller:
$$
C_d(z)=\frac{2.0107 (z-0.9605) (z^2 - 1.947z + 0.9481) (z^2 - 1.778z + 0.7941)}{(z-1) (z-0.9849) (z-0.9695) (z^2 - 1.777z + 0.7935)}
$$
with poles $1, \hspace{0.1cm} 0.9849,\hspace{0.1cm}  0.9695\hspace{0.1cm}   0.8887 \pm 0.0613i   0.8887 + 0.0613i$
As it can be seen, applying a mild dynamics separation rule, there is just one fast pole. Therefore, the parallel configuration will be:
$$
C_d (z)= 2.0107+b_1+b_2+b_3+b_{45}  \\
$$
being:
\begin{equation*}
    \begin{array} {lcl}
 b_1&=&\frac{0.1193}{z-1} \\
 b_2&=&\frac{-0.02817}{z-0.9849} \\
 b_3&=&\frac{0.001037}{z-0.9695} \\
 b_{45}&=&\frac{-0.0003399 z + 0.0001023}{ z^2 - 1.777 z + 0.7935}
 \end{array}
\end{equation*}

\begin{figure} \centering
\includegraphics[height=9cm, width=12cm]{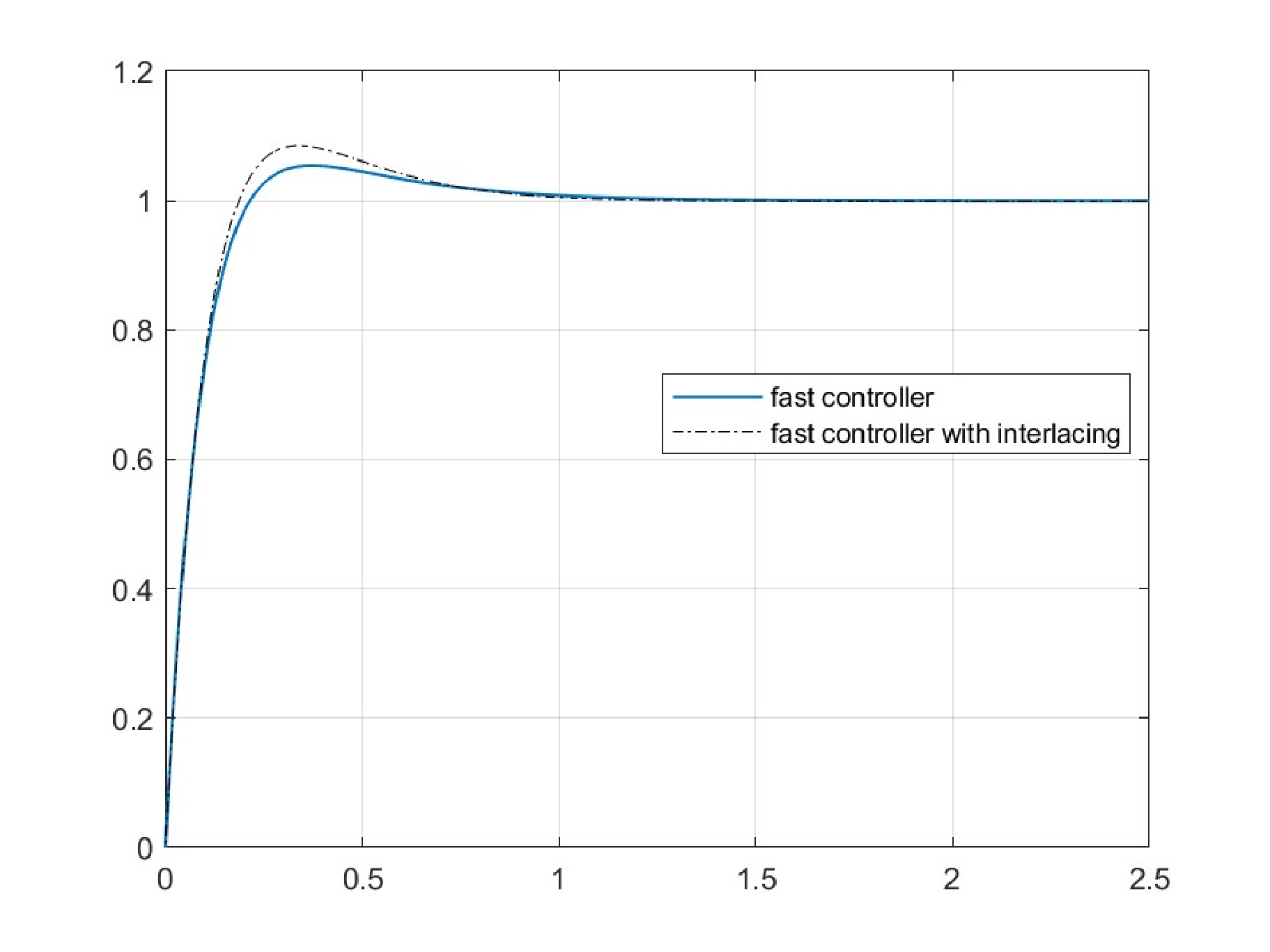}
\caption {Closed Loop Output Response with interlacing implementation} \label{old_blockorder}
\end{figure}

Using the explained procedure with:
\begin{equation*}
    \begin{array} {lcl}
 \mathcal{W}_1&=&z^2 +  z + 1 \\
\mathcal{W}_2&=&z^2 + 0.9849 z + 0.97 \\
\mathcal{W}_3&=&z^2 + 0.9695 z + 0.94 \\
 \end{array}
\end{equation*}
it is obtained the slow poles fast controller for implementing with fast input and slow output:
\begin{equation*}
    \begin{array} {lcl}
 b_1&=&\frac{0.358}{z^3-1} \\
 b_2&=&\frac{-0.08324}{z^3-0.9553} \\
 b_3&=&\frac{0.003016}{z^3-0.9114} \\
 b_{45}&=&\frac{-0.0003399 z + 0.0001023}{ z^2 - 1.777 z + 0.7935}
 \end{array}
\end{equation*}
it is noted that $z^3$ is the variable $z$ referred to $3T$.
Note that the blocks $C_{s,i}^T$ corresponds to the $block_i$ following the procedure in subsection \ref{Csi}.
Before the interlacing can be applied, it is necessary to test the slow single rate controller. In the case that this slow control would result in losing performance or even instability then it would be inadvisable to consider this kind of implementation.
The closed loop output time response with interlacing blocks implementation assuming the case (I-1,O-1) is depicted in figure \ref{old_blockorder}. 
%Different responses are obtained based on the different gains of the different blocks. It is observed that the response has a light chattering effect due to the fast switch of slow interlacing blocks.\\

\begin{figure} \centering
\includegraphics[height=9cm, width=12cm]{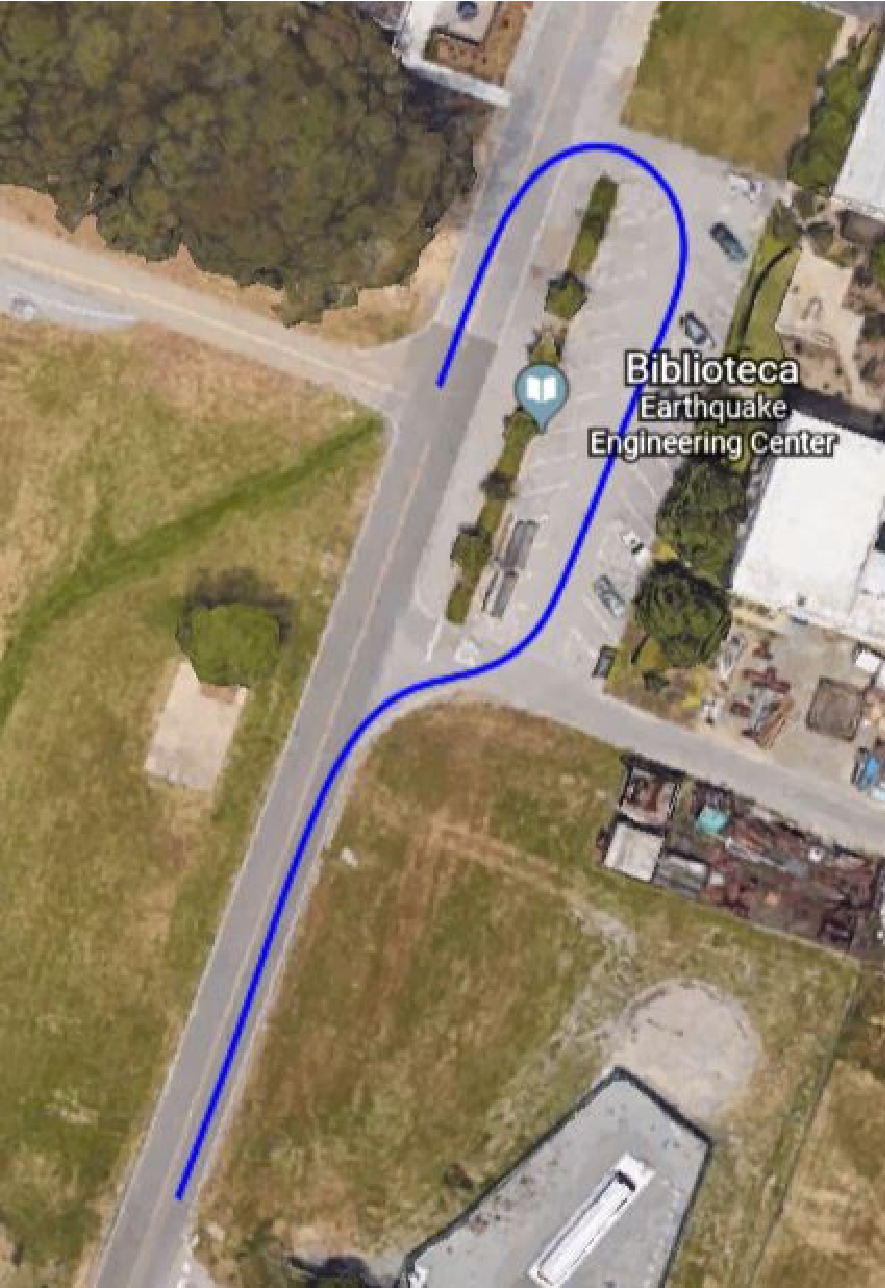}
\caption {U-turn test at Richmond Field (UCLB)} \label{Rfield}
\end{figure}

Finally, a car behavior test was performed with respect the Richmond Field
Station serving as a designated closed area (UC at Berkeley). In this case the u-turn scenario was selected (Figure \ref{Rfield}). The simulated test considered the way points and the velocities (with smooth steering) stored in a previous real test.

\begin{figure} \centering
\includegraphics[height=8cm, width=10cm]{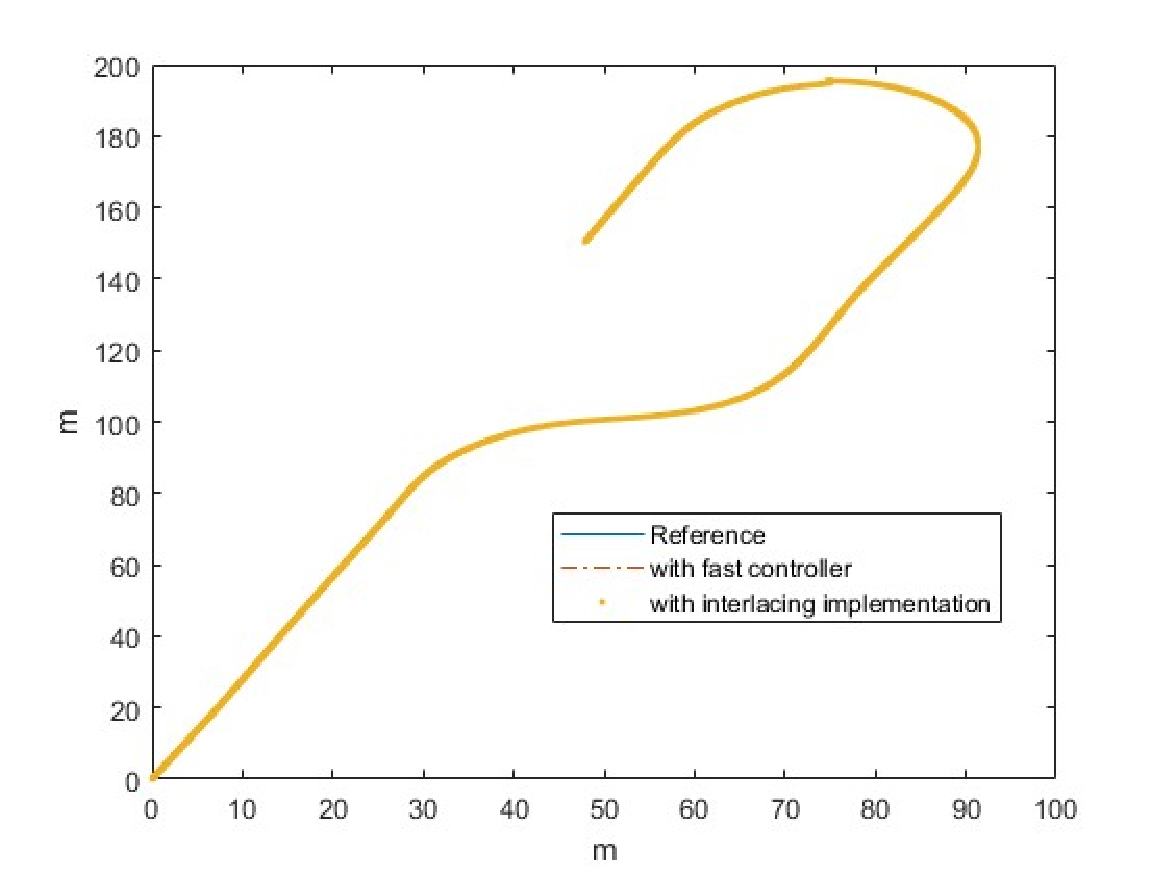}
\caption {Car u-turn trajectory with different implementations} \label{uturn}
\end{figure}

In figure \ref{uturn} is compared the car trajectories with the robust fast controller and its interlacing implementation. Both trajectories are overlapped in this figure and as it can be seen, it is difficult to appreciate the differences

%However, figure \ref{Inp_Out_dif} shows the different close loop outputs with different Input-Output implementation Strategies. It seems that the fast input sampling and Slow updating of slow interlacing block leads to a good response. The comparison of this last curve (I-1, O-1) with single rate (slow and fast) controllers is represented in figure \ref{SR_comparison}. In general, the responses using fast controller interlacing should be between both single rate (slow and fast) controllers. That is the reason why is important that the slow single rate response was suitable. Obviously the ideal solution would be a good approximation of the interlacing problem to the fast single rate case. The high slope around $20$ Rad/s explains the chattering.
%\begin{figure} \centering
%\includegraphics[height=8cm, width=10cm]{Compara_Input_Output12_34_5.eps}
%\caption {Closed Loop Output Response with different Strategies Input - Output} %\label{Inp_Out_dif}
%\end{figure}

%\begin{figure} \centering
%\includegraphics[height=8cm, width=10cm]{Compara_con_SRate.eps}
%\caption {Closed Loop Output Response Comparison with Single Rate Controllers} \label{SR_comparison}
%\end{figure}

\section{Conclusions}

This study has demonstrated that computational savings are possible in the lateral control of an autonomous vehicle using appropriate techniques for the implementation of controllers with minimal acceptable losses in lane keeping behavior. A formal model has been introduced for this type of system that will enable future studies. It has been applied to a real-world case study of an urban circuit at UC Berkeley with very good results.
\section*{Acknowledgements}
This work has been carried out thanks to the support of Project PID2023-151755OB-I00, funded by MICIU/AEI/10.13039/501100011033/ and by FEDER/EU.

\bibliographystyle{elsarticle-num}
%\section{References}

\bibliography{bookmv,AV_multirate}

\begin{thebibliography}{10}
\expandafter\ifx\csname url\endcsname\relax
  \def\url#1{\texttt{#1}}\fi
\expandafter\ifx\csname urlprefix\endcsname\relax\def\urlprefix{URL }\fi
\expandafter\ifx\csname href\endcsname\relax
  \def\href#1#2{#2} \def\path#1{#1}\fi

\bibitem{ducaju2020application}
J.~M.~S. Ducaju, C.~Tang, M.~Tomizuka, C.-Y. Chan, Application specific system identification for model-based control in self-driving cars, in: 2020 IEEE Intelligent Vehicles Symposium (IV), IEEE, 2020, pp. 384--390.

\bibitem{artunedo2024lateral}
A.~Artu{\~n}edo, M.~Moreno-Gonzalez, J.~Villagra, Lateral control for autonomous vehicles: A comparative evaluation, Annual reviews in control 57 (2024) 100910.

\bibitem{salt2021autonomous}
J.~M. Salt~Ducaj{\'u}, J.~J. Salt~Llobregat, {\'A}.~Cuenca, M.~Tomizuka, Autonomous ground vehicle lane-keeping lpv model-based control: Dual-rate state estimation and comparison of different real-time control strategies, Sensors 21~(4) (2021) 1531.

\bibitem{you2006active}
S.-S. You, H.-S. Choi, H.-S. Kim, T.-W. Lim, S.-K. Jeong, Active steering for intelligent vehicles using advanced control synthesis, International Journal of Vehicle Design 42~(3-4) (2006) 244--262.

\bibitem{bhattacharya2009control}
R.~Bhattacharya, G.~Balas, Control in computationally constrained environments, IEEE transactions on control systems technology 17~(3) (2009) 589--599.

\bibitem{wu2004performance}
S.~Wu, M.~Tomizuka, Performance and aliasing analysis of multi-rate digital controllers with interlacing, in: American Control Conference, 2004. Proceedings of the 2004, Vol.~4, IEEE, 2004, pp. 3514--3519.

\bibitem{ding2006multirate}
J.~Ding, F.~Marcassa, S.~Wu, M.~Tomizuka, Multirate control for computation saving, Control Systems Technology, IEEE Transactions on 14~(1) (2006) 165--169.

\bibitem{salt2014hard}
J.~Salt, M.~Tomizuka, Hard disk drive control by model based dual-rate controller. computation saving by interlacing, Mechatronics 24~(6) (2014) 691--700.

\bibitem{coulter1992implementation}
R.~C. Coulter, Implementation of the pure pursuit path tracking algorithm, Tech. rep. (1992).

\bibitem{kuwata2009real}
Y.~Kuwata, J.~Teo, G.~Fiore, S.~Karaman, E.~Frazzoli, J.~P. How, Real-time motion planning with applications to autonomous urban driving, IEEE Transactions on control systems technology 17~(5) (2009) 1105--1118.

\bibitem{ljungqvist2015motion}
O.~Ljungqvist, Motion planning and stabilization for a reversing truck and trailer system (2015).

\bibitem{buehler2009darpa}
M.~Buehler, K.~Iagnemma, S.~Singh, The DARPA urban challenge: autonomous vehicles in city traffic, Vol.~56, Springer Science \& Business Media, 2009.

\bibitem{rill2005vehicle}
G.~Rill, Vehicle dynamics, Fachhochschule Regensburg University of Applied Sciences (2005).

\bibitem{zhang2008development}
J.-Y. Zhang, J.-W. Kim, K.-B. Lee, Y.-B. Kim, Development of an active front steering (afs) system with qft control, International Journal of Automotive Technology 9~(6) (2008) 695--702.

\bibitem{wu2008precision}
S.-C. Wu, Precision control for high-density and cost-effective hard disk drives, University of California, Berkeley, 2008.

\bibitem{jia2008new}
Q.~Jia, A new method of multirate state feedback control with application to an hdd servo system, Mechatronics 18~(1) (2008) 13--20.

\bibitem{lopez2010two}
S.~L{\'o}pez-L{\'o}pez, A.~Sideris, J.~Yu, Two-stage design of multirate h8 optimal controllers, in: American Control Conference (ACC), 2010, IEEE, 2010, pp. 2647--2652.

\bibitem{arxiv_interlac}
J.~Salt, Interlacing in controllers implementation: Frequency analysis, https://doi.org/10.48550/arXiv.2510.20394 (2025).

\bibitem{lu1989least}
W.~Lu, D.~G. Fisher, Least-squares output estimation with multirate sampling, IEEE Transactions on Automatic Control 34~(6) (1989) 669--672.

\bibitem{francis1988stability}
B.~Francis, T.~Georgiou, Stability theory for linear time-invariant plants with periodic digital controllers, Automatic Control, IEEE Transactions on 33~(9) (1988) 820--832.

\end{thebibliography}

\end{document}